\documentclass[doublecol]{epl2} 

\title{Scaling of phononic transport with connectivity in amorphous solids }
\shorttitle{Phononic transport and connectivity} 
\author{Matthieu Wyart}

\institute{ Lewis-Sigler Institute, Princeton University, Princeton, NJ 08544;\\
Center for Soft Matter Research, New York University, NY 10003
}

\usepackage{graphicx}
\usepackage{bm}
\usepackage{psfrag}
\usepackage{subfigure}

\newcommand{\be}{\begin{equation}}
\newcommand{\ee}{\end{equation}}
\newcommand{\ba}{\begin{eqnarray}}
\newcommand{\ea}{\end{eqnarray}}

\abstract{
The effect of coordination on transport  is investigated theoretically using random networks of springs as model systems. An effective medium approximation
is made to compute the density of states of the vibrational modes, their energy diffusivity (a spectral measure of transport) and their spatial correlations as the network coordination $z$ is varied. Critical behaviors are obtained as $z\rightarrow z_c$ where these networks lose rigidity. A sharp cross-over from a  regime where modes are plane-wave-like toward a regime of extended but strongly-scattered modes  occurs at some frequency $\omega^*\sim z-z_c$, which does not correspond to the Ioffe-Regel criterion.  Above $\omega^*$ both the density of states and the diffusivity are nearly constant. These results agree remarkably with recent numerical observations of repulsive particles near the jamming threshold \cite{ning}.  The analysis further predicts that the length scale characterizing the correlation of displacements of the scattered modes decays as $1/\sqrt{\omega}$ with frequency, whereas for $\omega<<\omega^*$ Rayleigh scattering is found with a scattering length $l_s\sim (z-z_c)^3/\omega^4$. It is argued that this description applies to silica glass where it compares well with thermal conductivity data, and to transverse ultrasound propagation in granular matter. }

\begin{document}

\maketitle

The temperature dependence of the thermal conductivity in glasses above a few $K^0$ suggests a regime where the scattering of phonons is strong and nearly independent of frequency, as noticed by Kittel  \cite{kittel}. At these temperatures specific heat measurements indicate the presence of excess modes in comparison with the Debye model, the Boson Peak \cite{Phillipsbook}, whose amplitude correlates with other fundamental properties of glasses,  such as the fragility of the  super-cooled liquid \cite{sokolov}. Understanding the nature of the strongly scattered modes leading to an excess density of states  may therefore unravel some of the most mysterious properties of these materials. Several mean-field models have been proposed to explain the presence of such modes \cite{schi1,argentin,mayr} and their effects on transport. Nevertheless, there is yet no simple generic rules 
to predict from the microscopic structure of the glass the amplitude of the peak and the frequency at which transport becomes diffusive. Existing mean-field approaches do not explain why, for example,  silica glass presents one of the largest Boson peak, whereas it is almost inexistent for amorphous silicon, although both structure are tetravalent. The  spatial nature of the modes also remains unresolved.  Recently numerics have shown that the vibrational properties of repulsive, short-range particles display scaling behavior when decompressed toward the jamming transition \cite{J}. One observes a frequency scale $\omega^*$ which vanishes with decompression and marks a sharp cross-over toward a regime of strong scattering at larger frequency, where the modes diffusivity, which can be computed numerically via a Kubo formalism \cite{ning,vincenzo}, is nearly constant and small, in consistence with Kittel's conjecture,  and where the density of states displays a plateau. Close to the jamming threshold the cross-over toward strong scattering can occur at a wave-length infinitely larger than the particle size. Thanks to this separation of length scales this system is very suitable to study the spatial structure of the excess mode, and its dependence on the microscopic structure. A variational argument can explain the plateau in the vibrational spectrum, and shows that both the mean coordination and contact strain, averaged on a scale $l^*\sim1/(z-z_c)$ which diverges near threshold, control locally the onset of the excess modes\cite{matthieu1,matthieu2}- referred to as ``anomalous" to underline the difference of these excitations with plane waves. This length does not appear in the static structure, but  characterizes the heterogeneity of the response to a point force \cite{wouter}.  Nevertheless, this variational argument does not yield the spatial correlations of the displacements of the modes, nor their transport ability, or diffusivity. Interestingly, some aspects of the diffusivity can be estimated by making several assumptions on the spatial nature of the modes or their wave-vector decomposition \cite{vincenzo}. However, these assumptions are not trivial and lack theoretical justifications. 
  In this letter the spatial correlations of the modes and their diffusivity are computed by applying an effective medium approximation, which has been used in similar contexts \cite{thorpe,schi2} and in  electrical resistor networks \cite{kirpatrick}, to random networks of harmonic springs close to the minimal coordination where rigidity is lost and unjamming occurs. Results compare well with properties of sphere packing, and also with measurements of thermal conductivity in silica at high temperature.

I consider a system of point particles of unit mass $m\equiv1$, interacting via a lattice  of springs with coordination $z_0$ strictly larger than twice the spatial dimension $d$, where all bonds are assumed to be equivalent, such as in the triangular lattice. Springs are harmonic, unstretched,  of unit length $a\equiv 1$, and have a stiffness $k_0$, whose square root defines our unit frequency (thus $k_0=1$ in our units).  The energy expansion reads  $\delta E=\sum_{\langle ij\rangle} \frac{1}{2} [(\delta {\vec R_i}-\delta {\vec R_j})\cdot {\vec n_{ij}}]^2 +o(\delta R^2)$, where the sum is on all bonds $ij$, whose direction is along the unit vector $\vec n_{ij}$. Springs are randomly removed with a probability $1-p$, such that the coordination of the remaining network is $z=p z_0$. The static elastic properties and the presence of vibrational modes of zero frequency  of such a disordered lattice have been  studied previously \cite{thorpe}.   Here we extend the analysis to compute vibrational properties in the mechanically stable phase. 

The random network is modeled as an effective medium of spring stiffness $k_M(\omega)$.  To determine $k_M(\omega)$, one considers the ``defectuous" network where all bonds have a stiffness $k_M(\omega)$, except for one bond $(ij)=(12)$ at the origin, whose stiffness $k_{12}$ is distributed according to the distribution $P(x)=p\delta(x=1)+(1-p)\delta(x=0)$ of the disordered system. Next one considers a homogeneous oscillatory stress  applied on this system. If the bond $(12)$ were identical to the other bonds, the applied stress would lead to an affine oscillatory response $\delta{\vec R}^a_i(t)\equiv\delta{\vec R}^a_i(\omega) \exp(i\omega t)$, $i=1...N$. Such a response is not a solution in general, as if it is imposed on this network, an unbalanced oscillatory force dipole of amplitude $f_{12}(\omega)=(k_{12}-k_M(\omega)) e(\omega)$ appears on the bond $12$, where $e(\omega)=(\delta{\vec R}_1(\omega)-\delta{\vec R}_2(\omega))\cdot {\vec n_{12}}$ is the contact strain. The solution is obtained by adding the response of this dipole to the affine displacement. This causes  an extra strain $e^*(\omega)$ to appear in the bond (12).  $k_M(\omega)$ is determined self-consistently by imposing that the effective medium respond with the correct average strain to an imposed stress, implying that $\langle e^*(\omega)\rangle=0$, where $\langle X\rangle=\int X(k_{12}) P(k_{12}) dk_{12}$ . In general this requirement leads to a complex solution for $k_M(\omega)$ , reflecting the broken symmetry of translation in the disordered network and the presence of scattering.

The extra strain $e^*(\omega)$ can be expressed in terms of the Green function of the effective homogeneous system \cite{kirpatrick,webman,thorpe}, as is now sketched. Newton's equation for an oscillatory external force field $\delta{\vec F}_i(t)=\delta{\vec F}_i(\omega) \exp(i\omega t)$ leads to $\sum_j A_{ij}(\omega) \cdot \delta {\vec R}_j(\omega)=\delta{\vec F}_i(\omega)$, where: 
\be
\label{1}
A_{ij}(\omega)= \delta_{ij}(-\omega^2 +\sum_{\langle l\rangle} k_M  {\vec n_{il}}\otimes {\vec n_{il}}) -\delta_{\langle ij\rangle} k_M  {\vec n_{ij}}\otimes {\vec n_{ij}}
\ee
In Eq.(\ref{1})  $\otimes$ is the tensor product, the sum is made on all neighbors $l$ of $i$, and $\delta_{\langle ij\rangle}=1$ if $i$ and $j$ are neighbors, and zero otherwise. In these notations $A_{ij}$ is a tensor of rank $d$. We denote by $A(\omega)$ the $dN\times dN$ linear application relating displacements to forces field. Its inverse is the Green function $G=A^{-1}$. $G_{ij}$ is the tensor of rank $d$ characterizing the displacement of particle $i$ in response to a force applied in $j$. In  the ''defectuous'' network the additional displacement field $\delta {\vec R_i^*}$, $i=1...N$, generated by the force dipole in the bond $12$ satisfies $\forall i$: 
\begin{eqnarray*}
\label{2}
&&\sum_j A_{ij}  \delta {\vec R}_j^*- (\delta_{i1}-\delta_{i2})(k_{12}-k_M){\vec n_{12}}\otimes{\vec n_{12}}\cdot (\delta {\vec R}_1^*-\delta {\vec R}_2^*) \\
&&=-{\vec n_{12}}(k_M(\omega)-k_{12}) e(\omega) (\delta_{i1}-\delta_{i2})  
\end{eqnarray*}
After some algebra one gets:
\begin{eqnarray*}
\label{3}
\lefteqn{e^*(\omega)=(\delta{\vec R^*}_1(\omega)-\delta{\vec R^*}_2(\omega))\cdot {\vec n_{12}}} \\
&=& \frac{2 e(\omega)(k_M-k_{12}){\vec n_{12}}\cdot (G_{11}(\omega)-G_{12}(\omega))\cdot {\vec n_{12}}}{1+2(k_{12}-k_M){\vec n_{12}}\cdot (G_{11}(\omega)-G_{12}(\omega))\cdot {\vec n_{12}}}
\end{eqnarray*}
This expression can be simplified  using that $\sum_j  A_{ij} G_{ji}$is the unit tensor of rank $d$, Eq.(\ref{1}), and the assumption that all bonds are equivalent \cite{kirpatrick}. One obtains if $i$ and $j$ are neighbors that ${\vec n_{ij}}\cdot (G_{ii}(\omega)-G_{ij}(\omega))\cdot {\vec n_{ij}}=d(1+ {\vec n_{ij}}\cdot G_{ii}(\omega)\cdot {\vec n_{ij}})/(k_M z_0)$. Applied in Eq.(\ref{3}), and requiring $\langle e_0(\omega)\rangle =0$ leads to:
\be
\label{4}
\langle \frac{k_{12}-k_M}{k_M+(k_{12}-k_M)\frac{2d}{z_0}(1+\omega^2  {\vec n_{12}}\cdot G_{11}(\omega)\cdot {\vec n_{12}})}\rangle=0
\ee
Note that in this expression, ${\vec n_{12}}$ can be replaced by any contact direction  ${\vec n_{ij}}$ and $ G_{11}(\omega)$ by $G_{ii}(\omega)$. 

Taking the static limit  $\omega\rightarrow 0$ in Eq.(\ref{4}) and using $pz_0\equiv z$, one gets $k_M=(z-2d)/(z-2d p)$. Thus as  found by Thorpe and collaborators \cite{thorpe}, the effective stiffness of the network, and therefore all the elastic constants, vanish linearly as $z\rightarrow 2d$,  as found numerically in generic elastic networks \cite{wouter2}. $z=2d$ corresponds to the Maxwell bound where rigidity is lost  \cite{max}, such network are called isostatic. Note that for sphere packing, the bulk modulus does not vanish near the isostatic point, as explained in \cite{these}. For such purely repulsive systems I expect the forthcoming result on the Rayleigh scattering to hold for transverse modes only. 

\begin{figure}       
  \centering       
  \vspace{0.5cm}                  
   \rotatebox{-0}{\resizebox{8.9cm}{!}{\includegraphics{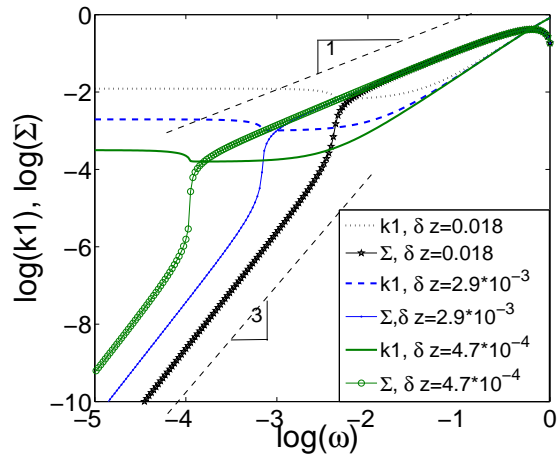}}}
   \caption{\label{f0}Logarithmic plot  of the real and imaginary part of the effective stiffness $k_M\equiv k_1-i\Sigma$ {\it vs} frequency for several coordination numbers $\delta z\equiv z-z_c$.  }
   \vspace{0 cm}
\end{figure}

The Green function can be computed numerically for any lattice, and used to solve the self-consistent Eq.(\ref{4}). In what follows we focus on the low-frequency scaling properties near this critical point. One expects such properties to be universal and not to depend on the details of the lattice. We shall therefore consider a simplified model, where the lattice elasticity is characterized by a single elastic constant $B$, which is taken equal to $k_M$. For the isotropic lattices, such as the hexagonal lattice, two elastic constant are required to describe elasticity. The analysis below is readily extended to this case, and leads to the same scaling behavior. We consider for the Green function:
\be
\label{5}
G_{i1}(\omega)=  \int_0^{k_{max}}\frac{d^dk}{(2\pi)^d}\sum_\alpha \frac{e^{i{\vec k}\cdot{\vec r_i}}}{ k_M(\omega) k^2-\omega^2} \vec{ k}_\alpha\otimes  \vec{ k}_\alpha
\ee
where the Debye cut-off $k_{max}$ is a constant in our units, and is set to one in what follows. ${\vec r_i}$ is the vector going from particle 1 to $i$, and the sum is made on the different polarization $\vec{k}_\alpha$. All spatial directions being equivalent, one obtains:
\begin{eqnarray}
\hbox{ trace}(G_{11}(\omega))/d={\vec n_{12}}\cdot G_{11}(\omega)\cdot {\vec n_{12}}  \\
= \int_0^{1} \frac{d^dk}{(2\pi)^d} \frac{1}{ k_M(\omega) k^2-\omega^2} {} \label{6}
\end{eqnarray}
 Eqs.(\ref{4},\ref{6}) are solved numerically in the three-dimensional case, using  $z_0=12$.  The results for the real and imaginary part of  $k_M\equiv k_1(\omega)-i\Sigma(\omega)$ are shown in Fig.(\ref{f0}) for three different coordination. Qualitatively these results can be obtained by approximating the real part of the integral of Eq.(\ref{6}) by  its large $k$ contribution (obtained by setting $\omega=0$),
and its imaginary part by the singular value of the integral, plus its large $k$ contribution. This is approximately valid for $k_1\geq \Sigma$.
Expanding Eq.(\ref{4}) at low frequency, one finds the following schematic set of equations (where numerical pre-factors are omitted):
\ba
\label{10}
k_1=\delta z-\frac{\omega^2 k_1}{k_1^2+\Sigma^2}\\
\label{11}
\Sigma=\frac{\omega^3}{k_1^{3/2}}+ \frac{\Sigma\omega^2}{k_1^2+\Sigma^2}
\ea
where $\delta z\equiv z-z_c$ characterizes the distance from threshold. These schematic equations lead to a sharp cross-over at $\omega^*=\delta z/2$. This scaling is checked in inset of Fig.(\ref{2}). For smaller frequencies, $k_1\approx \delta z$, as expected from the static analysis, and $\Sigma\approx \omega^3/\delta z^{3/2}<<k_1$. At $\omega^*$, $k_1$ drops by  some numerical factor, and $\Sigma$ increases by a factor of order $1/\sqrt{\delta z}$. For larger frequencies, $\Sigma >> k_1$, and the (middle) term of  Eq.(\ref{11}) corresponding to the singular value of the integral  is not present. This leads to $\Sigma \sim \omega$.

\begin{figure}       
  \centering       
  \vspace{0.5cm}                  
   \rotatebox{+0}{\resizebox{6.9cm}{!}{\includegraphics{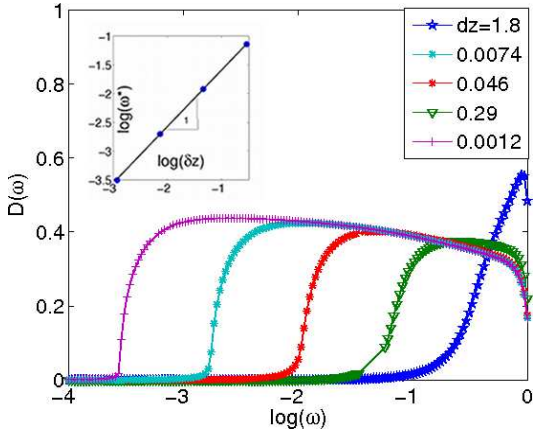}}}
   \caption{\label{f1}Density of states {\it vs} frequency $\omega$ for five coordination numbers.  Inset: Log-log plot of $\omega^*$, defined as  $D(\omega^*)=0.2$,  {\it vs} $\delta z$. The line has a slope 1. }
   \vspace{-0.cm}
\end{figure}

The density of states $D(\omega)$ of the effective medium is readily extracted using $D(\omega)=(2\omega/\pi) \hbox{Im}(\hbox{trace}(G_{11}(\omega))$,
and is shown in Fig.(\ref{f1}). The cross over at $\omega^*$ corresponds to a sharp transition from a Debye behavior $D(\omega)\sim \delta z^{-3/2}\omega^2$ at low frequency toward a plateau at higher frequency, and therefore correspond to the boson peak. The same cross-over is  observed numerically for elastic particles near jamming \cite{J,matthieu2}.

\begin{figure}       
  \centering       
  \vspace{0.5cm}                  
  \rotatebox{-0}{\resizebox{6.9cm}{!}{\includegraphics{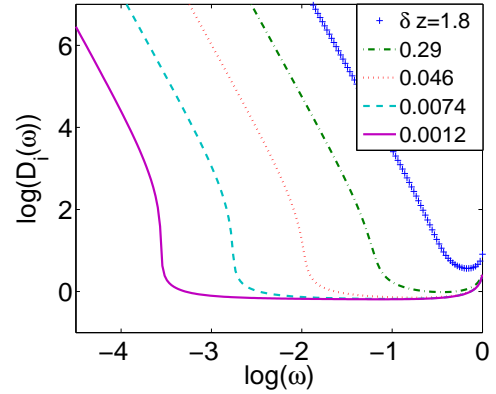}}}
  \caption{\label{f2} Log-log plot of the diffusivity $D_i(\omega)$ {\it vs} $\omega$ for several coordination numbers, as indicated in the legend. 
  }
  \vspace{0 cm}
\end{figure}

We next consider the spatial Green function at frequency $\omega$. For large distances $r_i>>1$ one finds:
\be
\label{7}
\hbox{trace}(G_{i1}(\omega))= 3\frac{e^{i\omega r_i/v(\omega)}e^{-r_i/l_s(\omega)}}{4\pi r_i k_M}
\ee
 $l_s(\omega)$ and $v(\omega)$ can be viewed as the scattering length and the  velocity of the corresponding vibrations, and their expression is:
\ba
l_s(\omega)&=&\frac{|k_M|}{\omega |\hbox{Im}(k_M^{1/2})|}\\
v(\omega)&=&\frac{|k_M|}{\hbox{Re}(k_M^{1/2})}\label{22}
\ea
Transport at frequency $\omega$ is characterized by the diffusivity $D_i(\omega)$, whose approximation $ D_i(\omega)= v(\omega)l_s(\omega)/3$ is shown in Fig(\ref{f2}).
This approximation is expected to be accurate  for weak-scattering ($\omega<<\omega^*$) and crude but qualitatively correct for the strongly-scattered modes  ($\omega>>\omega^*$) as long as they are extended, since it enforces the scaling requirement that the diffusivity is the product of a characteristic speed times a characteristic length. A more accurate computation of the diffusivity would require to go beyond the single bond effective medium investigated here.   

For $\omega<<\omega^*$,  Rayleigh scattering occurs: $D_i(\omega)\sim \delta z^{7/2}/\omega^4$.  At $\omega\sim\omega^*$, there is a sudden decay of the scattering length from $1/\delta z$ toward $1/\sqrt \delta z$  which is of the order  of the wavelength $\lambda\equiv v/\omega$. This sudden decay of the scattering length at  $\omega^*$ implies that this cross-over cannot  be identified from the quasi-plane wave regime $(\omega<<\omega^*)$ using the Ioffe-Regel criterion $l_s=\lambda$. 
  For $\omega>>\omega^*$ there is a flat plateau in the diffusivity, again in very good agreement with numerical studies of transport of jammed elastic particles \cite{ning,vincenzo}. This plateau stems from the compensation of two effects: the decrease of the scattering length $l_s(\omega)\sim \omega^{-1/2}$, and the inverse increase of the velocity $v(\omega)\sim \omega^{1/2}$. Note that at the high frequency end of the spectrum a localization edge, rather insensitive  to coordination, is observed numerically \cite{ning} and is not captured by this analysis.

The spatial properties of the strongly scattered normal modes $|e_\omega\rangle$ are deduced from the relation $\langle {\vec e}_\omega(1)\otimes  {\vec e}_\omega(i)\rangle \propto \hbox{Im}(G_{1i}(\omega))\omega/D(\omega)$. Thus Eq.(\ref{7}) implies that at $\omega^*$, the correlation length of the displacement  is of order $\delta z^{-1/2}$, and decreases as $\omega^{-1/2}$ for higher frequencies. 
Scaling arguments for why anomalous modes near $\omega^*$ are characterized by a length of order  $\delta z^{-1/2}$ -which is the value of the wavelength just below $\omega^*$ \cite{Leo2}-  have been given by Vitelli et al. \cite{vincenzo}. 

\begin{figure}       
  \centering       
  \vspace{0.5cm}                  
  \rotatebox{0}{\resizebox{8.5cm}{!}{\includegraphics{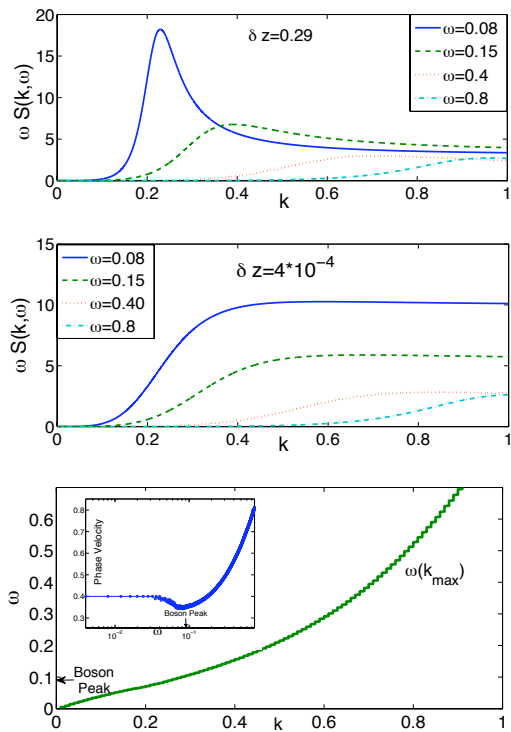}}}
    \vspace{-0.8 cm}
  \caption{ \label{f4} Dynamical structure factor $S(k,\omega)$ times $\omega$ {\it vs}  wave vector $k$ for different frequencies indicated in legend. Upper panel: $\delta z=0.29$. Middle panel: $\delta z= 4*10^{-4}$. No peaks are apparent. Lower panel: Dispersion relation $k_{max}(\omega)$ as defined in the text for $\delta z=0.29$. Inset:  phase velocity $\omega/k_{max}$ {\it vs} frequency $\omega$ for $\delta z=0.29$. The velocity displays a minimum close to the Boson peak frequency $\omega_{BP}=0.09$. 
  }
  \vspace{-0.5 cm}
\end{figure}

 Experimentally the dynamical structure factor $S(k,\omega)$ is accessible. In Fig.(\ref{f4}) the quantity  Im($k^2/(k_Mk^2-\omega^2))$, proportional to $\omega S(k,\omega)$ for a harmonic dynamics, is plotted. At threshold, no  peaks are found in the structure factor, which simply consists of a gap at low frequency, which ends approximately when  $\omega^4\sim \Sigma^2 k^4$ leading to  $k\sim \omega^{1/2}$, before reaching a plateau. At larger coordination,  well defined-peaks for  $\omega<\omega^*$ appear in the spectrum, corresponding to weakly-scattered plane waves. At larger frequencies the peak erodes, and the structure factor eventually resembles what is found at the critical point. The intensity gap at small frequency and the presence of a large background of large spatial frequencies have been observed above the Boson peak peak  in various systems such as silica \cite{ruocco} and sphere packing \cite{Leo2}.  
 
I define a dispersion relation by considering  the maximum $k_{max}(\omega)$ of $S(k,\omega)$ at fixed $\omega$. As shown in the lower panel of Fig.(\ref{f4}), this dispersion relations is linear at low frequency, and presents an inflection point near the Boson peak, before increasing as $\omega\sim k^2$. This inflection point leads to a drop in the velocity $\omega/k_{max}$ at the Boson peak frequency as shown in the inset of Fig.(\ref{f4}), a rather subtle effect recently observed in Lennard-Jones simulations  \cite{mossa}.


{\it Applications to amorphous solids}: Numerical observations show that when friction is added, the scaling relations valid for the vibrational spectrum of purely repulsive particles still hold \cite{wim}. Thus the results on transport derived here are expected to hold in granular matter as well, at least in the limit where dissipative effects  \cite{jia0} do not dominate sound propagation. In small systems of grains,  low-frequency ultrasounds propagate ballistically, whereas propagation is diffusive at high-frequency \cite{Jia}. According to the present analysis the cross-over wavelength between these regimes should decrease with coordination. The coordination depends on the system preparation and the friction coefficient, and can be measured independently by comparing transverse and longitudinal speed of sound \cite{bruno,jacob,roux}.

\begin{figure}       
  \centering       
  \vspace{0.5cm}                  
  \rotatebox{-0}{\resizebox{6.9cm}{!}{\includegraphics{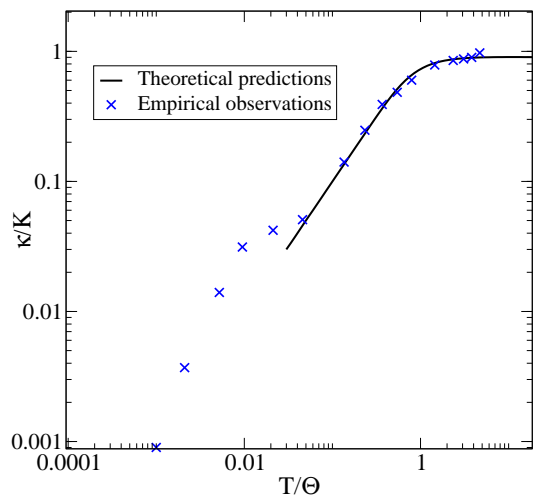}}}
  \caption{\label{silica} Log-log plot of the thermal conductivity in silica glass against temperature in arbitrary units. Observations taken from Freeman and Anderson \cite{freeman} appear in crosses. Full line: prediction for a flat spectrum of anomalous modes with constant diffusivity.  
  }
  \vspace{0 cm}
\end{figure}

Covalent glasses are also well suited to study the effect of coordination on transport.  Practically important examples are tetravalent covalent networks where the joints linking tetrahedra have a small bending force constant such as in silica glass (where the Si-O-Si angle is easily deformable) or germanium oxide. If these joints are are assumed to be freely rotating and only the other covalent interactions are taken into account (the so-called RUM model \cite{silica1}), such networks turns out to be isostatic and display a flat density of states up to zero frequency  \cite{silica2}, in consistence with the present analysis and previous arguments \cite{these,matsi}. When the small but finite bending force constant among tetrahedra is taken into account, the plateau of anomalous modes in the spectrum is shifted and appears above 1Thz  \cite{these,matsi}, corresponding to the Boson peak frequency. Most of the spectrum thus consists of anomalous modes which must display a constant diffusivity according to the present analysis. Assuming a constant density of states and diffusivity up to some maximal frequency $\omega_{max}$ enables to compute the thermal conductivity $\kappa(T)=\int_{0}^{\infty} {D}(\omega) C(\omega) {D_i}(\omega) d\omega\propto \int_{0}^{\omega_{max}}  C(\omega) d\omega$ where $C(\omega)\propto (\omega/T)^2 \exp({\bar h}\omega/kT)/( \exp({\bar h}\omega/kT)-1)^2$ is the heat capacity of the modes at frequency $\omega$. Fig.(\ref{silica}) shows that such an estimate captures accurately the thermal conductivity for several decades above its plateau. In this high-temperature estimate, I have neglected the contribution of plane-wave like modes at low frequency. This is a good approximation since the contribution of these modes can be read at lower temperature where only those modes affect transport and is small. The plane waves contribution is of order of the thermal conductivity plateau where the cross-over occurs, which is significantly smaller than the high temperature conductivity.  This analysis thus supports that anomalous modes are responsible for transport at room temperature.
Note that in tetravalent structures with strong bending force constant at the joints, such as in amorphous silicon where the same covalent bonds characterize the joints and the tetrahedra themselves, the effective coordination is high, and a Boson peak is hardly detectable, as expected.
The present results are expected to apply in simple molecular liquids where long-range interactions are present as well \cite{these,xu}. 

%

{\it Comparison with the variational argument of \cite{matthieu1}}:
In \cite{matthieu1} a variational argument guessing the nature of the modes leading to a plateau in the density of states was able to explain the abrupt change in this quantity at some frequency $\omega^*\sim \delta z$.
The corresponding anomalous modes are not built by softly distorting translational modes like plane waves are. The view that the Boson peak is a cross-over between two distinct types of excitations is reflected by the sharp cross-overs in the density of states and the transport properties at $\omega^*$ found here, supporting that new objects suddenly appear in the spectrum at this frequency. Although anomalous modes also can be characterized by a length scale and a velocity, their spatial structure  qualitatively differs from plane waves as appears in their dynamical structure factor, flat  with a gap at small frequencies rather than peak-like as for plane waves.

If the  variational procedure of \cite{matthieu1} and the effective medium theory developed here lead to consistent results, these approaches are also complementary. The variational argument suggested that  anomalous modes would lead to poor transport \cite{these,vincenzo}, but did not enable to compute their dynamical structure factor  explicitly nor the correlation length $l_s\sim 1/\sqrt {\delta z}$ of their displacement field, as was done here. One the other hand, the present approach does not yield the length scale $l^*\sim 1/\delta z$ on which fluctuations in the microscopic structure affect the anomalous modes \cite{matthieu1} and the response to a local perturbation \cite{wouter}. This may be a generic limitation of mean field approaches.

{\it Comparison with previous mean field approaches}:
Previous mean-field analysis of the Boson peak, see e.g. \cite{schi1,mayr,argentin},  have also connected this phenomenon to the presence of an elastic instability. Nevertheless the nature of the instability in those cases is singularly different from the present study, for example they find that the density of states does vanish at zero frequency at the instability threshold. This discrepancy  underlines that two parameters, and two frequency scales, must be included to describe excess modes even in simple amorphous solids.  In particular, the present analysis has not included the effect of compression on the vibrational spectrum, or equivalently  the presence of a finite strain in the contacts.   This is  a good approximation for covalent glasses and granular matter at limited pressures, but compression plays an important role e.g. for colloids near the glass transition where it even determines the microscopic structure of the glass \cite{brito}. For any fixed coordination $z>2d$, increasing compression shifts anomalous modes toward lower frequencies and eventually leads to an elastic instability as shown in \cite{matthieu2}, for which the density of states does  vanish at zero frequency. Near this instability, the spectrum displays two decoupled characteristic frequency: a Boson peak frequency where the first anomalous modes appear and a higher frequency $\omega^*$ above which the density of states displays a plateau \cite{matthieu2}, and where the present analysis is expected to hold.  These two characteristic frequencies are identical only when pressure is negligible, in general they can be different, reflecting that two parameters, coordination and pressure, strongly affect the vibrational spectrum. Previous mean-field analysis \cite{schi1,mayr,argentin} are presumably well suited to describe transport properties at low-frequency near the instability driven by compression. Including compression to the present effective medium approach will assess this view, and may provide a  unified description of  transport  applicable to a broad range of amorphous solids.

It is a pleasure to thank V. Vitelli for discussions stimulating this work, A. Liu, S. Nagel and N. Xu for discussions, and D. Huse and M. Muller for comments on the manuscript.

\end{document}